\documentclass[twoside,twocolumn,prl,final,superscriptaddress,aps,a4paper,10pt]{revtex4-1}
\usepackage{graphicx} 
\usepackage{dcolumn} 
\usepackage{bm} 
\usepackage{amsmath}
\usepackage{amssymb}
\usepackage{amsfonts}
\usepackage{bbm}
\usepackage{ae}
\usepackage{epsfig}
\usepackage{multibib}
\usepackage[usenames,dvipsnames]{color}
\usepackage{xspace}
\usepackage[paperwidth=210mm,paperheight=297mm,centering,hmargin=2cm,vmargin=2.65cm]{geometry}
\usepackage{balance}

\usepackage[latin1]{inputenc}
\usepackage[T1]{fontenc} 

\newcommand{\gu}{\ensuremath{\ket{g\!\uparrow}}}
\newcommand{\gd}{\ensuremath{\ket{g\!\downarrow}}}
\newcommand{\eu}{\ensuremath{\ket{e\!\uparrow}}}
\newcommand{\ed}{\ensuremath{\ket{e\!\downarrow}}}
\newcommand{\egp}{\ensuremath{\ket{eg^+}}\xspace}
\newcommand{\egm}{\ensuremath{\ket{eg^-}}\xspace}

\newcommand{\Vex}{\ensuremath{V_{\text{ex}}}}

\newcommand{\ket}[1]{\ensuremath{\left|#1\right\rangle}}
\newcommand{\sz}{$^1$S$_0$}
\newcommand{\pz}{$^3$P$_0$ }

\newcommand{\fs}[1]{\textcolor{Black}{#1}}

\begin{document}

\title{Observation of two-orbital spin-exchange interactions with ultracold SU($N$)-symmetric fermions} 

\author{F. Scazza}\affiliation{Ludwig-Maximilians-Universität, Schellingstraße 4, 80799 München, Germany}\affiliation{Max-Planck-Institut für Quantenoptik, Hans-Kopfermann-Straße 1, 85748 Garching, Germany}
\author{C. Hofrichter}\affiliation{Ludwig-Maximilians-Universität, Schellingstraße 4, 80799 München, Germany}\affiliation{Max-Planck-Institut für Quantenoptik, Hans-Kopfermann-Straße 1, 85748 Garching, Germany}
\author{M. Höfer}\affiliation{Ludwig-Maximilians-Universität, Schellingstraße 4, 80799 München, Germany}\affiliation{Max-Planck-Institut für Quantenoptik, Hans-Kopfermann-Straße 1, 85748 Garching, Germany}
\author{P. C. De Groot}\affiliation{Ludwig-Maximilians-Universität, Schellingstraße 4, 80799 München, Germany}\affiliation{Max-Planck-Institut für Quantenoptik, Hans-Kopfermann-Straße 1, 85748 Garching, Germany}
\author{I. Bloch}\affiliation{Ludwig-Maximilians-Universität, Schellingstraße 4, 80799 München, Germany}\affiliation{Max-Planck-Institut für Quantenoptik, Hans-Kopfermann-Straße 1, 85748 Garching, Germany}
\author{S. Fölling}\email{simon.foelling@lmu.de}\affiliation{Ludwig-Maximilians-Universität, Schellingstraße 4, 80799 München, Germany}\affiliation{Max-Planck-Institut für Quantenoptik, Hans-Kopfermann-Straße 1, 85748 Garching, Germany}

\date{\today}

\begin{abstract}
\noindent
Spin-exchanging interactions govern the properties of strongly correlated electron systems such as many magnetic materials. When orbital degrees of freedom are present, spin exchange between different orbitals often dominates, leading to the Kondo effect, heavy fermion behaviour or magnetic ordering. Ultracold ytterbium or alkaline-earth ensembles have attracted much recent interest as model systems for these effects, with two (meta-) stable electronic configurations representing independent orbitals.
We report the observation of spin-exchanging contact interactions in a two-orbital SU($N$)-symmetric quantum gas realized with fermionic $^{173}$Yb.  We find strong inter-orbital spin-exchange by spectroscopic characterization of all interaction channels and demonstrate SU($N=6$) symmetry within our measurement precision. The spin-exchange process is also directly observed through the dynamic equilibration of spin imbalances between ensembles in separate orbitals. The realization of an SU($N$)-symmetric two-orbital Hubbard Hamiltonian opens the route to quantum simulations with extended symmetries and with orbital magnetic interactions, such as the Kondo lattice model.
\end{abstract}

\maketitle
Ultracold gases of alkaline-earth-like atoms have recently been the focus of increasing theoretical \cite{Gorshkov2010a} and experimental \cite{Sugawa2011, Taie2012-Pomeranchuk, Martin2013, Pagano2014} efforts, owing to the versatile internal structure which makes them attractive candidates for the study of quantum many-body physics. 
One of the remarkable properties of alkaline-earth-like atoms is the existence of a long-lived metastable excited electronic state, owing to the separate singlet and triplet electronic spin manifolds. The correspondingly low linewidth of the associated optical ``clock'' transitions has enabled the realization of the most precise atomic clocks \cite{Hinkley2013, Bloom2014}.
More recently, the existence of two stable electronic states has also inspired many proposals for the implementation of previously inaccessible fundamental many-body systems \cite{Gorshkov2010a, Foss-Feig2010a, Gerbier2010-Gauge} and quantum information processing schemes \cite{Daley2008-QuantComp, Gorshkov2009-FewQubit}. These proposals are motivated by the very different ways in which atoms in the two states interact both with light and with other atoms. This is in direct analogy to electrons in a crystal lattice occupying two different orbitals, an essential part of fundamental condensed-matter models such as the Kondo lattice \cite{RudermanKittel, Tsunetsugu1997, Foss-Feig2010a, Foss-Feig2010b} and the Kugel-Khomskii model \cite{KugelKhomski, Gorshkov2010a}. Such models describe a large class of materials \cite{Tokura2000, ColemanBook2007, gegenwart2008}, and rely on the coexistence of electrons in two orbitals which are coupled by a spin-exchange interaction. Using different electronic states to simulate the orbital degree of freedom has been proposed as a solution to the challenge of experimentally implementing these Hamiltonians to address the still many open questions.

A second striking property of alkaline-earth-like atoms is the large decoupling between electronic and nuclear degrees of freedom which is expected for states with total electronic angular momentum $J=0$ \cite{Boyd2006, Gorshkov2010a}. This induces an SU($N$)-symmetric situation, where collisional properties are independent of the nuclear spin orientation. Systems possessing such high-dimensional symmetries are predicted to exhibit a variety of still unexplored many-body phases \cite{Honerkamp2004, Hermele2009, Cazalilla2009, Hermele2011} or could be used to simulate high-energy non-Abelian gauge theories \cite{Banerjee2013-GaugeTheories}. In our case, using $^{173}$Yb and choosing the metastable state $\ket{e}=\,^3$P$_0$ as the second orbital in addition to the ground state $\ket{g}=\,^1$S$_0$, the realization of a SU($N=6$)-symmetric two-orbital model is expected. The existence of spin-exchanging coupling terms, which lie at the heart of orbital magnetism, as well as the stability of the ensuing ensemble depend on specific properties of the interactions between $\ket{g}$ and $\ket{e}$ atoms, and have previously been unknown.

Interactions between ultracold Yb atoms in the electronic ground state are well characterized for most isotopes \cite{Kitagawa2008}, and estimates of the relevant interaction processes occurring in atomic clock experiments with fermionic $^{171}$Yb ($I=1/2$) were given \cite{Ludlow2011Shift, Lemke2011p-wave}. More recently, the interaction strength between the ground state and the $^3$P$_2$ state of bosonic $^{174}$Yb ($I=0$) has been determined in an optical lattice \cite{Yamaguchi2008-inelastic, Kato2013-ControlInteraction}.
In the present work, we characterize the specific properties of the \sz\,-\pz interaction channels of fermionic \textsuperscript{173}Yb ($I=5/2$) and directly reveal their SU($N$)-symmetric nature. Most importantly, we experimentally demonstrate strong spin-exchanging interactions between atoms in the two different electronic states.
The interaction channels are characterized by measuring interaction clock shifts, for different $m_F$ state combinations and magnetic field strengths. The magnetic field causes a state mixing, leading to a novel tunability of one of the Hubbard model interaction parameters.
Spin-exchanging collisions between the two electronic orbitals are observed directly in our system via the spin imbalance evolution in large ensembles.
Such spin-exchange interactions were observed for localized pairs of bosons in different motional states of a single lattice site \cite{Anderlini07}. Excited motional states however are not stable in the presence of strong tunnelling, a necessary component for realizing many-body systems such as the Kondo lattice model \cite{Gorshkov2010a}. In contrast, the long lifetime of the metastable state in Yb is not compromised by a high mobility of one orbital, as long as the strongly inelastic collisions between two \ket{e}-state atoms can be avoided.

\section{Two-orbital SU($N$)-symmetric interactions}
\noindent
The interactions between ground state atoms in an SU($N$)-symmetric degenerate $^{173}$Yb gas can be fully characterized by a single \textit{s}-wave scattering length $a_{gg} = 199.4\,a_0$ \cite{Kitagawa2008}: the scattering length is independent of the nuclear spin orientation, preventing therefore spin-changing collisions \cite{Stellmer2011}. 
Moreover, \textit{p}-wave collisions can be neglected, as they are strongly suppressed in the ultracold temperature regime.
\begin{figure}[t]
\includegraphics[width=\columnwidth]{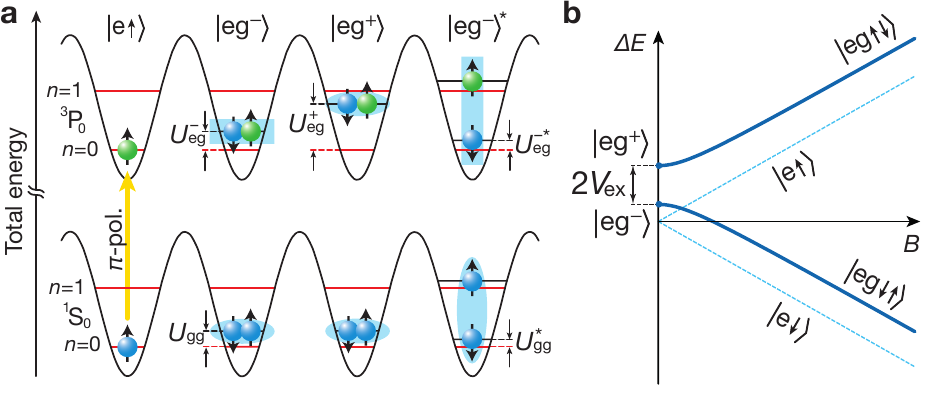}
\caption{\textbf{Two-orbital interacting states of fermions in a lattice.} (\textbf{a}) One- and two-particle states for interacting atoms on a lattice site with both orbital (\ket{g}: blue and \ket{e}: green) and nuclear spin (\ket{\uparrow},\ket{\downarrow}) degrees of freedom. The corresponding energy levels are also illustrated. A blue ellipse (rectangle) indicates a  nuclear spin singlet (triplet). (\textbf{b}) Sketch of the magnetic field dependence of the energy for an interacting atom pair on a lattice site (blue solid line). Energies of one-particle excited states are also drawn for comparison (dashed line). The zero here corresponds to the resonance of the clock transition for singly-occupied sites at zero magnetic field. For large Zeeman shifts ($|\Delta_B| \gg |\Vex|$) the two-particle eigenstates are the $\ket{eg\!\uparrow\downarrow}$ and $\ket{eg\!\downarrow\uparrow}$ states, as defined in the text.}
\label{fig:fig1ab}
\end{figure}

Introducing an additional internal degree of freedom, represented here by the electronic state, extends the system to a two-orbital description, where more collision channels become available. 
With the two electronic orbitals $\ket{g}$ and $\ket{e}$, four \textit{s}-wave scattering lengths are sufficient to describe collision processes for all nuclear spin combinations due to the SU($N$) symmetry. 
The scattering lengths $a_{gg}$ and $a_{ee}$ apply to atom pairs in the states $\ket{gg}$ and $\ket{ee}$, where both atoms are in the same electronic states $\ket g$ and $\ket e$, respectively. In addition, interactions between atoms in different electronic states, which are responsible for spin exchange between orbitals, are described by two parameters $a_{eg}^{\pm}$ for the symmetric and antisymmetric combinations $(\ket{eg} \pm \ket{ge})/\sqrt{2}$. 
We consider the specific case of one $g$ and one $e$ atom in the lowest vibrational state of the same lattice site, with two different nuclear spin states  $\ket{\uparrow}$ and $\ket{\downarrow}$, usually chosen to have opposite $m_F$ values for simplicity. 
As the fermionic statistics enforces the total state to be antisymmetric, two states are possible, which we denote as $\ket{eg^+}=(\ket{eg} + \ket{ge})/\sqrt{2} \otimes \ket{s}$ and $\ket{eg^-}=(\ket{eg} - \ket{ge})/\sqrt{2} \otimes \ket{t}$, where \ket{s} and \ket{t} are the nuclear spin singlet and triplet states. The relevant states are illustrated in Fig.~\ref{fig:fig1ab}(a).
Given the scattering lengths introduced above, the on-site Hubbard interaction strength for the different states can be written as:
\begin{align}
\label{eq:Ueg}
U_{X} = \frac{4 \pi \hbar^2}{m}\,a_{X}\int \! d^3r\, w_{a}^2(\textbf{r}) w_{b}^2(\textbf{r}) \:,
\end{align}
where $X=gg,\,ee,\,eg^+,\,eg^-$. Here, $m$ is the atomic mass and $w_{a,b}(\textbf{r})$ are the Wannier functions of the two atoms. The above expression is valid in the regime of sufficiently weak interactions ($a_X \ll a_{\text{ho}}$, with $a_{\text{ho}}$ being the on-site harmonic oscillator length), where the two-particle wavefunction can be expressed as a product of single-particle Wannier functions. In all computations, we assume negligible shifts of all elastic interactions from inelastic contributions.

In the presence of an external magnetic field $\textbf{B}$, an additional contribution arises due to the differential Zeeman shift between the \ket{g} and the \ket{e} states with a given $m_F$ \cite{Boyd2007}, which causes the linear magnetic shift of single-atom $\ket{g,m_F}\rightarrow\ket{e,m_F}$ clock transitions. The field introduces a coupling between $\ket{eg^+}$ and $\ket{eg^-}$ and the Hamiltonian for the two-atom system with a single electronic excitation in the $\left\{\ket{eg^+},\ket{eg^-}\right\}$ basis then has the form of a coupled two-level system: 
\begin{align}
H_{eg}=\left( \begin{array}{cc}
U_{eg}^+ & \Delta_B \\
\Delta_B & U_{eg}^- \end{array} \right)\:,
\label{eq: Hamiltonian}
\end{align}
where $\Delta_B=\delta g \,  m_F \mu_B B$ is the differential Zeeman shift. Here, $\delta g$ is the differential nuclear Land\'{e} $g$-factor \cite{Boyd2007}, $m_F$ is the nuclear spin projection along the field and $\mu_B$ is the Bohr magneton.  
Diagonalization results in two eigenenergy branches, represented in Fig.~\ref{fig:fig1ab}(b):
\begin{equation}
\label{eq:eigenvalues}
E_{1,2} = V \pm \sqrt{\Vex^2 + \Delta_B^2}\,,
\end{equation}
where $V=\left(U_{eg}^+ + U_{eg}^-\right)/2$ and $\Vex = \left(U_{eg}^+ - U_{eg}^-\right)/2$.
The zero-field eigenstates are then $\ket{eg^+}$ and $\ket{eg^-}$, whereas for high fields their superpositions form the eigenbasis: $\ket{eg\!\uparrow\downarrow}=(\eu\!\gd-\gd\!\eu)/\sqrt{2}\,$, $\ket{eg\!\downarrow\uparrow}=(\ed\!\gu-\gu\!\ed)/\sqrt{2}$.

\section{Interaction spectroscopy in a 3D lattice}
\noindent
We perform our experiments employing a two-component degenerate Fermi gas of \textsuperscript{173}Yb, which is loaded into a deep state-independent optical lattice potential and probed with $\pi$-polarized clock excitation light. 
The experimental methods for preparation and detection are described in the Methods section, with additional technical details in the Supplementary Information.
\begin{figure}[!t]
\includegraphics[width=\columnwidth]{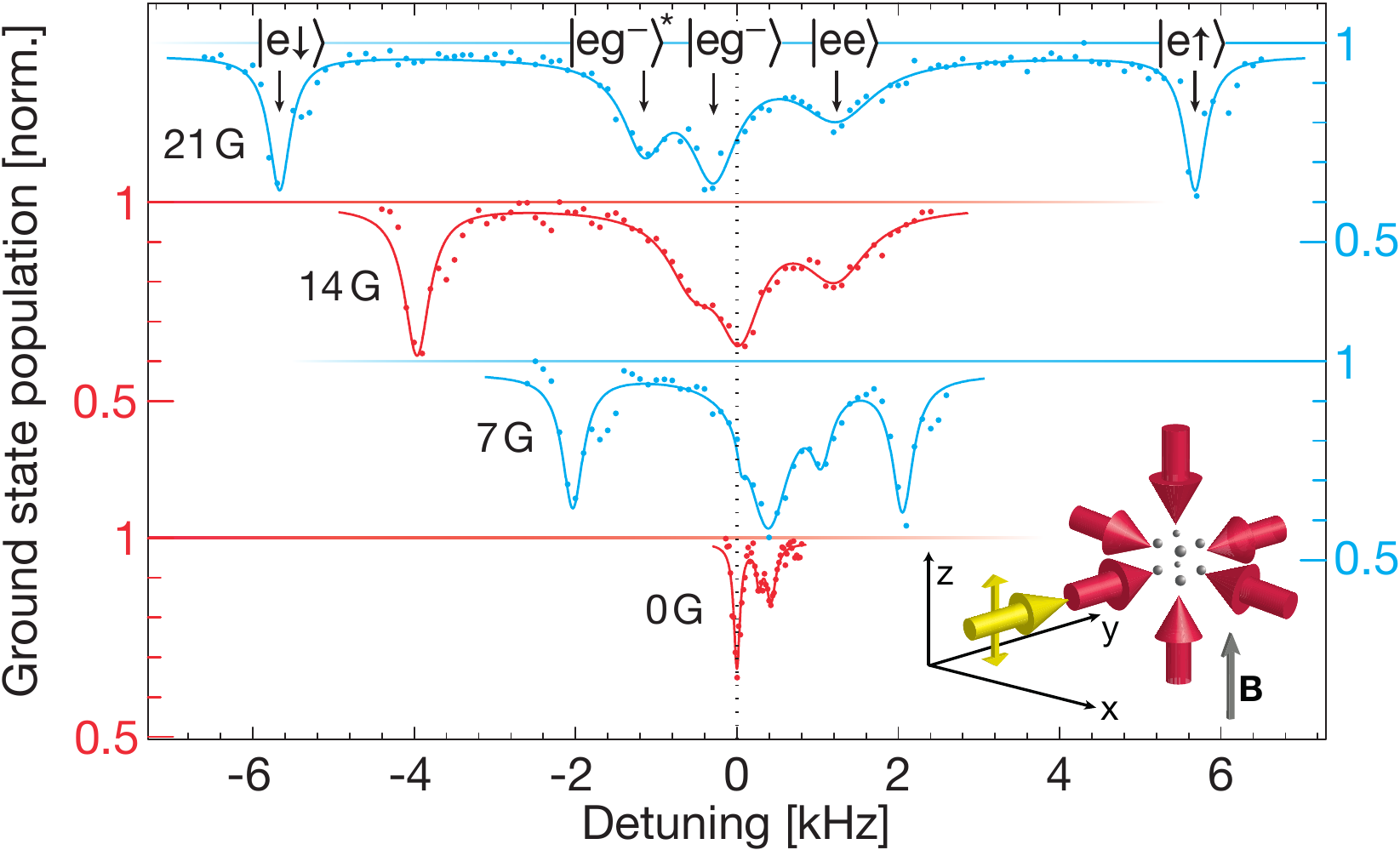}
\caption{\textbf{Clock transition spectroscopy of a two-component Fermi gas in a 3D lattice}. The absorption signals for the $m_F=\pm5/2$ spin mixture are shown, displaying resonances corresponding to different final states. Solid lines are multiple-Lorentzian fits to determine the resonance positions.  Singly-occupied lattice site resonances shift proportionally to the magnetic field due to the differential Zeeman shift between \ket{g} and \ket{e} states, whereas doubly-occupied site resonance shifts are dependent upon the final state of the excitation. An excitation light intensity of 7.5\,mW/cm$^2$ was used in these experimental runs, except for the 0\,G data where a reduced light intensity of 2.3\,mW/cm$^2$ was used to improve the resolution. The experimental configuration is schematically represented in the lower-right corner.}
\label{fig:egpPlusSpectro}
\end{figure}
\begin{figure}[!t]
\includegraphics[width=\columnwidth]{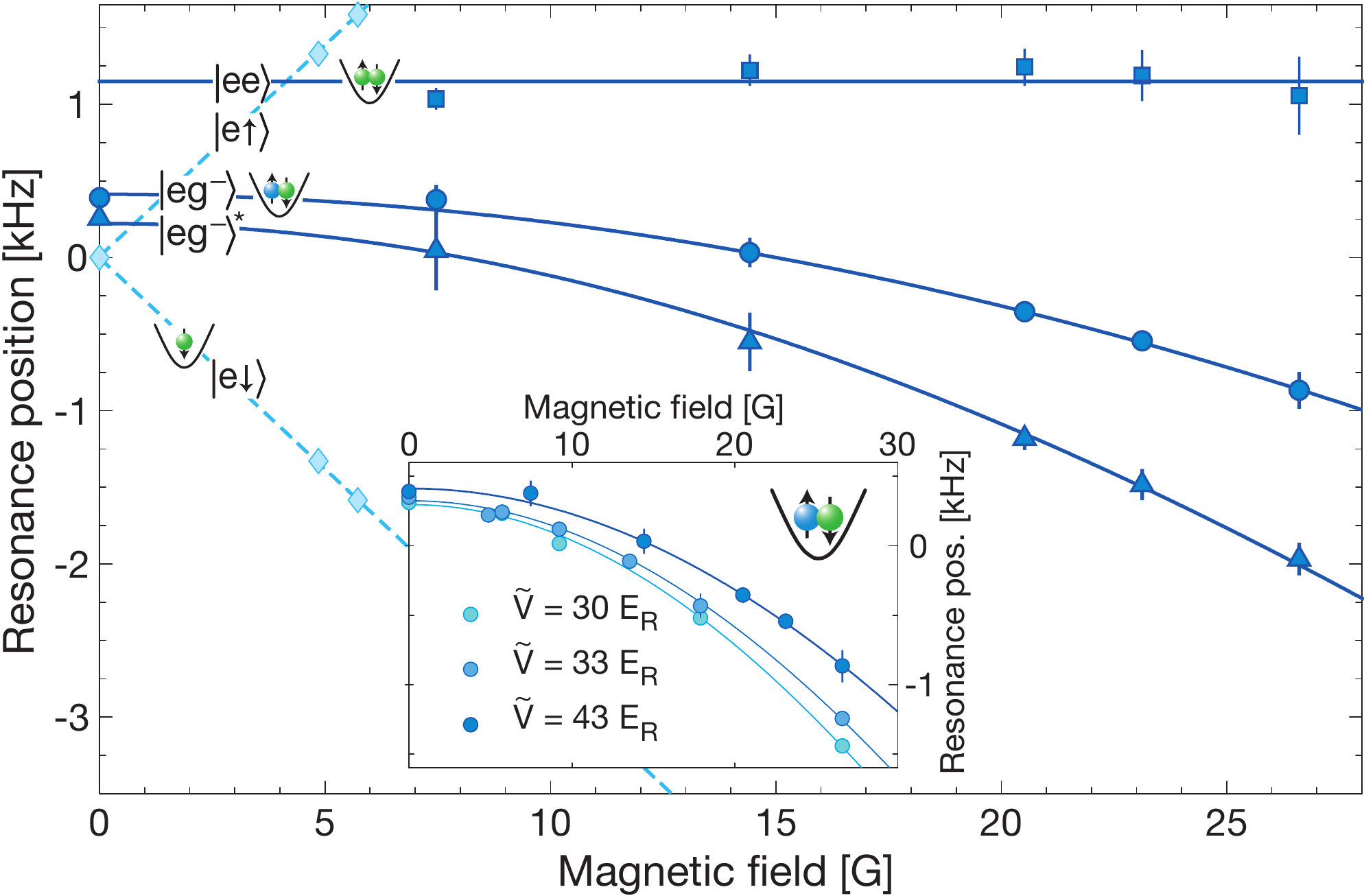}
\caption{\textbf{Magnetic field dependence of clock transition frequencies in the presence of two-orbital on-site interactions.} Resonance positions in an $m_F=\pm5/2$ spin mixture are shown for $\tilde{V}=43 E_r$, where $\tilde{V}=(V_x V_y V_z)^{1/3}$ is the mean lattice depth in units of recoil energy of the lattice. Squares mark transitions to \ket{ee}, circles to \fs{$\ket{eg^-}$}, triangles to \fs{$\ket{eg^-}^*$} and diamonds to $\eu$ and $\ed$ states. Resonance shifts of \fs{$\ket{eg^-}$} and \fs{$\ket{eg^-}^*$} states are fitted according to the model in Eq.~(\ref{eq:eigenvalues}) (solid blue lines). The doubly-excited $\ket{ee}$ state resonance position does not shift with magnetic field, as the two differential Zeeman shifts cancel each other. Error bars are $95\%$ confidence intervals of resonance position fits. Eq.~(\ref{eq:eigenvalues}) fit results for this lattice configuration are: $\Vex=h \times (22.2 \pm 1.0)$\,kHz and $V=U_{gg}+ h \times (22.6 \pm 1.9)$\,kHz. Inset: the \fs{$\ket{eg^-}$} resonance shifts are plotted at varying magnetic field for three different lattice depths.}
\label{fig:egPlusBFieldDependence}
\end{figure}

By loading a two-spin mixture into a 3D lattice, we obtain an average filling between $\bar{n}=1$ and $\bar{n}=2$ per lattice site and a temperature above the spin-1/2 Mott insulating transition.
The interaction shift of the \sz$\rightarrow$\pz transition is then a direct measurement of the difference between $U_{gg}$ and \fs{$U_{eg}^-$}, as illustrated in Fig.~\ref{fig:fig1ab}(a). 
Without magnetic field or other symmetry-breaking effects, the initial state $\ket{gg}\otimes \ket{s}$ is \fs{predominantly} coupled to the \fs{$\ket{eg^-}$} state. 
An increasing magnetic field coupling changes the energy of the lower eigenstate of Eq.~(\ref{eq: Hamiltonian}), and probing this dependence allows for determination of $\Vex$ through Eq.~(\ref{eq:eigenvalues}).
Scans of the probe laser detuning displaying various absorption lines in a $m_F=\pm5/2$ spin mixture are shown in Fig.~\ref{fig:egpPlusSpectro}.
The singly-occupied lattice sites shift linearly with magnetic field strength with a slope of $ m_F \times 110(5)$\,Hz/G. The frequency offset is calibrated using the centre of these two resonances.
Three other absorption lines are clearly visible, associated to the excitation of doubly-occupied lattice sites.
In Fig.~\ref{fig:egPlusBFieldDependence} the positions of the different resonances are plotted as a function of the magnetic field for a $m_F=\pm5/2$ spin mixture. 
We identify the strongest transition without linear Zeeman shift as the lower eigenstate of Eq.~(\ref{eq:eigenvalues}) and fit the resonance positions with $V$ and $\Vex$ as free parameters (see Supplementary Information). In this way we obtain the values of $U_{eg}^+ - U_{gg}$ and $U_{eg}^- - U_{gg}$ for each different lattice depth and spin mixture.
From Eq.~(\ref{eq:Ueg}) and taking identical lowest-band Wannier functions for the two atoms, we calculate for each fit result of \fs{$U_{eg}^- - U_{gg}$} the corresponding scattering length and obtain as mean value: \fs{$\Delta a_{eg}^- = a_{eg}^- - a_{gg} = (20.1 \pm 2.0)\,a_0$}. 
However, fit estimates of \fs{$U_{eg}^+$} exceed the energy gap to the first excited band of the lattice, causing Eq.~(\ref{eq:Ueg}) to be inaccurate for computing the scattering length. A more suitable model including more bands is therefore required, which will be discussed below.
Consistent values of \fs{$\Delta a_{eg}^-$} are obtained from spectroscopic measurements of several distinct spin mixtures, thereby demonstrating the SU($N$) symmetry of this scattering channel down to our experimental uncertainty of 0.9\% of \fs{$a_{eg}^- = 219.5\,a_0$}. The reported value of \fs{$\Delta a_{eg}^-$} comprises all our measurements, including the case of a balanced six-spin mixture. The magnetic field dependent shift of the transition is measured for all employed spin mixtures as well, yielding estimates of $\Vex$ in good agreement with each other and validating therefore also the SU($N$) symmetry of the \fs{symmetric} scattering channel (see Supplementary Information).  
The \fs{two-particle nature of the transition to the $\ket{eg^-}$ state} is also confirmed by the fact that we find the Rabi frequency to be larger than on the bare atom resonance by the expected factor of $\sqrt{2}$.

An additional absorption line is detected close to the \fs{$\ket{eg^-}$} line, which we associate with the final state \fs{$\ket{eg^-}^*$}, schematically represented in Fig.~\ref{fig:fig1ab}(a).
This transition is analogous to the \fs{$\ket{gg}\rightarrow\ket{eg^-}$} transition with both states having one excitation to the first vibrational state.
Fitting the data and using Eq.~(\ref{eq:Ueg}), in this case with different Wannier functions for the two atoms, yields \fs{$\Delta a_{eg}^- = (22.7 \pm 7.3)\,a_0$}, consistent with the value obtained from the \fs{$\ket{eg^-}$} resonance. 
We find this line to have a significant weight when large atom numbers are used, suggesting that the loading to the 3D lattice is not fully adiabatic and the second band starts to become populated.
Finally, we identify the resonance which exhibits no magnetic field shift with the transition to the $\ket{ee}$ state, corresponding to a detuned two-photon transition with \fs{$\ket{eg^-}$} as the intermediate state. Its coupling strength is therefore dependent on the separation to the \fs{$\ket{eg^-}$} line. Applying Eq.~(\ref{eq:Ueg}) to data separately for each adopted spin combination, we obtain a mean value $\Delta a_{ee} = a_{ee} - a_{gg} = (106.8 \pm 10.4) \,a_0$.
A direct excitation resonance to the \fs{$\ket{eg^+}$} state was not identified. Its strength is expected to be suppressed by a reduced wavefunction overlap and, for our low magnetic fields, \fs{owing to the fact that the Clebsch-Gordan coefficients of the $\pi$-transitions for the two $m_F$-states have opposite sign}. In addition, its location is expected close to the band excitation energy, in direct proximity to the blue Raman sidebands of the \fs{$\ket{eg^-}$} and single atom transitions. 

The values of $\Vex$ obtained by making use of the simple model in Eq.~(\ref{eq:eigenvalues}) are as large as $\Vex \simeq h\times 22\,$kHz, corresponding to \fs{$U_{eg}^+ \gtrsim  h\times 44\,$}kHz, in comparison to a typical lattice band gap of only $\simeq h\times 25$\,kHz, depending on the lattice configuration. In such a regime the strong interaction couples the excited bands of the lattice \cite{busch98, Will2010a}, and the full band structure along with proper regularization of the interaction potential needs to be taken into account. As a consequence, the single-band model in Eqs.~(\ref{eq:Ueg})-(\ref{eq:eigenvalues}) breaks down: the on-site pair wavefunction is modified by the strong \fs{$U_{eg}^+$} interaction.
For increasing interaction strengths, the wavefunction overlap between the two atoms is reduced compared to the lowest band Wannier function, up to the point where the atom pair ``fermionizes'': for an infinite scattering length the interaction integral vanishes, and the interaction energy \fs{$U_{eg}^+ = U_{eg}^- + 2 \Vex$} saturates to the first excited band energy \cite{paredes04,Zurn2012-Fermionization}. 
Using Eq.~(\ref{eq:Ueg}) in this regime \fs{($a_{eg}^+ > a_{\text{ho}}$)} for a given value of $U$ therefore strongly underestimates the scattering length \cite{busch98}. 
Applying Eq.~(\ref{eq:Ueg}) yields \fs{$\Delta a_{eg}^+ = a_{eg}^+ - a_{gg} = (1.97\pm 0.19)\times 10^3\,a_0$}, which should be then taken only as a lower estimate.  A numerical diagonalization including four bands, using an unregularized delta-potential was carried out, for which we find a larger value \fs{$\Delta a_{eg}^+ \simeq 4 \times 10^3\,a_0$} to best reproduce the data. Even assuming full fermionization of the on-site wave function, the calculated magnetic field dependence of the lower eigenenergy branch is still close to the measured data (see Supplementary Information).
In the regime \fs{$a_{eg}^- > a_\text{ho}$}, the scattering length dependence of the signal is low and very precise modelling of the system including proper regularization, the complete internal structure and the full anisotropic lattice appears necessary to constrain the scattering length further when using on-site interaction data.
\section{Direct observation of spin-exchange dynamics}
\noindent
It can be easily seen that the interaction energy difference between the $\ket{eg^+}$ and $\ket{eg^-}$ state causes spin exchange when considering as initial state $\ket{eg\!\uparrow\downarrow}$, which is a superposition of $\ket{eg^+}$ and $\ket{eg^-}$. The energy difference between these two states then causes oscillations at zero magnetic field with a frequency equal to $\left| \Vex \right|/h$, between the $\ket{eg\!\uparrow\downarrow}$ and $\ket{eg\!\downarrow\uparrow}$ states, in analogy to the vibrational states in \cite{Anderlini07}.
In order to generate the necessary superposition of $\ket{eg^+}$ and $\ket{eg^-}$ eigenstates at zero field, large optical couplings $\Omega$ on the order of $V_{ex}/\hbar$ would be required. Alternatively, excitation in a large magnetic field and subsequent non-adiabatic switching to zero field would require a switching time scale $\ll h/|\Vex|$. In our setup, neither option can be achieved owing to the large measured $\Vex$.

\begin{figure}[thb]
\includegraphics[width=\columnwidth]{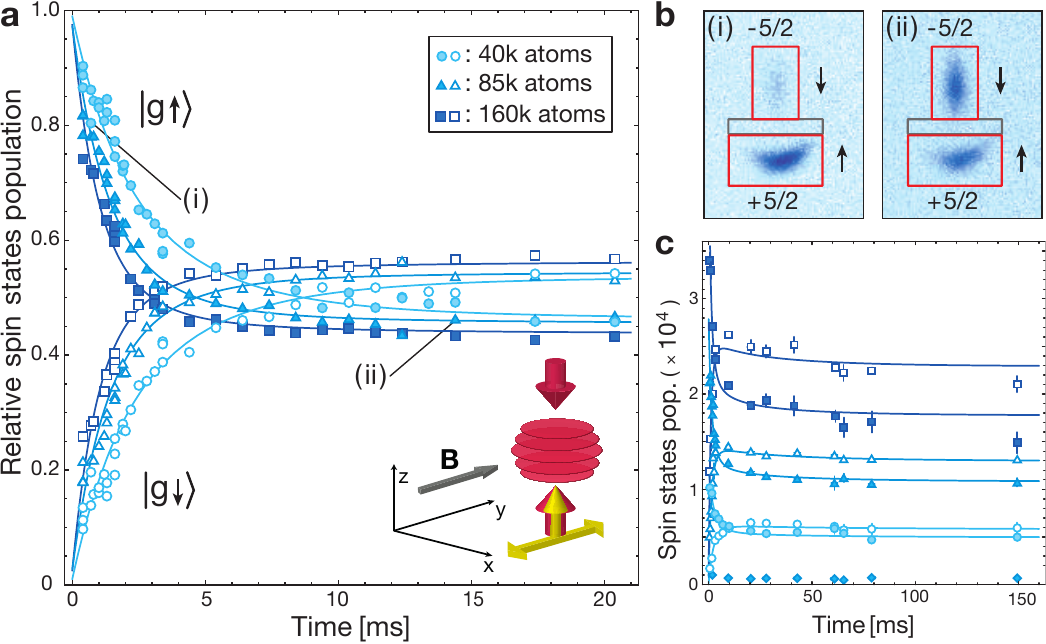}
\caption{
\textbf{Spin-exchange dynamics between \textbf{$\ket{g}$} and \textbf{$\ket{e}$} atoms in 2D ensembles.} (\textbf{a}) The relative populations of the two ground states $\gu$ ($m_F=+5/2)$ and $\gd$ ($m_F=-5/2)$ in a 1\,G bias field are shown for variable hold times, after preparation in an approximately equal mixture of $\gu$ and $\ed$. The fitted curves are obtained with a two-body rate equation model that includes inelastic losses. The modified experimental configuration is schematically represented in the lower-right corner. (\textbf{b}) Examples of spin-resolved absorption images: 0.4\,ms hold time (i) and 14\,ms hold time (ii). Atom counting regions for the different spin states are shown: the solid red rectangles correspond to atoms occupying the $m_F=\pm5/2$ states, the grey rectangle corresponds to atoms occupying other intermediate spin states. (\textbf{c}) Long-time evolution of the spin distribution: a weak loss of ground state atoms due to inelastic collisions is visible. Intermediate spin state populations are shown as filled diamonds for the $N=0.85\times10^5$ atoms case. No re-population of other spin states is detected. Error bars denote the standard deviation of the mean obtained by typically averaging 2-3 measurement points.}
\label{fig:spinExchangeEvolution}
\end{figure}

In order to still directly observe dynamics driven by the exchange coupling, we use a different configuration. Ensembles of atoms are loaded into 2D traps formed by a single, vertical standing wave and the excitation beam is also directed along the vertical axis (see Fig.~\ref{fig:spinExchangeEvolution}(a)).
We prepare a mixture of $\gu$ and $\ed$ atoms by magnetically splitting the transitions to the $\eu$ and $\ed$ states in a degenerate two-component ($m_F=\pm5/2$) Fermi gas, and addressing only the $\ket{\downarrow}$ state with a $\pi$-pulse. The transition is blue-shifted by typically $\simeq 0.5$\,kHz with respect to the same transition in a spin-polarized gas due to the interaction shift. 
Subsequently, the magnetic field is rapidly reduced in $200\,\mu$s to a low value and the atoms are held in the vertical lattice for a variable time. The spin distribution is measured using absorption imaging with an optical Stern-Gerlach scheme.

As expected for a situation with large exchange interactions, we observe a fast redistribution of the two spin components as soon as the bias field is being reduced toward its small stationary value of $1\,$G.
The equilibration takes place without populating any other spin state, indicating genuine spin-exchange between the two electronic states without spin-changing collisions, further confirming the SU($N$)-symmetric character of this process.
For long equilibration times a ground state spin distribution with a stationary ratio according to the spin distribution prior to excitation is observed.

We fit the spin state population evolution with a typical two-body collision rate equation model, with a resonant $e$--$g$ spin-exchange collision term, a slow inelastic $e$--$g$ collisional loss term and fast inelastic $e\!\uparrow$ -- $e\!\downarrow$ collisions (see Supplementary Information).
The use of this type of rate equation system, more commonly used for collision dynamics in classical gases, is an approximation motivated by the known presence of strong $e$--$e$ losses which suppress the build-up of coherence between $\ket{eg\!\uparrow\downarrow}$ and $\ket{eg\!\downarrow\uparrow}$ pair states. The model also does not account for non-classical many-body correlations during the evolution.

Fits to the data obtained for a set of three different atom numbers are shown in Fig.~\ref{fig:spinExchangeEvolution}. Assuming a simple finite-temperature Fermi gas density model for each of the vertical lattice sites (see Supplementary Information) results in a spin-exchange rate coefficient $\gamma_{ex} = (1.2 \pm 0.2) \times 10^{-11}$\,cm$^3$/s. 
Inelastic $e$--$e$ collisions are separately measured and a loss rate coefficient $\beta_{ee} = (2.2 \pm 0.6) \times 10^{-11}$\,cm$^3$/s is obtained based on the same density model used for the spin-exchange rate. This value is comparable to previous measurements in $^{171}$Yb \cite{Ludlow2011Shift}.
Inelastic $g$--$e$ state losses are also characterized by other independent measurements (see Methods and Supplementary Information). 
These measurements and the data in Fig.~\ref{fig:spinExchangeEvolution}(c) yield a loss rate coefficient $\beta_{eg} = (3.9 \pm 1.6) \times 10^{-13}$\,cm$^3$/s for our density model. This aggregate value includes both symmetric and antisymmetric loss channels. The uncertainties given for the three reported rate coefficients are due to uncertainties in the determination of the in-plane densities combined with fit uncertainties.
The spin distribution evolutions obtained by rate equation fits are displayed as solid lines in Fig.~\ref{fig:spinExchangeEvolution}(a)-(c).

The specific loss rate of \fs{$\ket{eg^-}$} atom pairs is additionally characterized in a deep 3D lattice configuration, yielding a two-body loss rate coefficient \fs{$\beta_{eg}^- < 3 \times 10^{-15}$\,cm$^3$/s}, corresponding to an imaginary part of the scattering length \fs{$\left| \eta_{eg}^- \right| < 0.006$\,$a_0$} (see Supplementary Information). This implies that the aggregate $g$--$e$ loss process measured in the 1D lattice is largely due to inelastic collisions in the \fs{$eg^+$} channel.

\section{Discussion}
\noindent
The measured rate coefficients in the $e$--$g$ mixture can not be reliably compared quantitatively to the much more precise values measured on-site, without further modelling which incorporates many-body correlations. Nevertheless, the direct comparison of the combined strength of the \fs{$eg^+$} loss channel and the spin-exchange processes both measured in the same configuration confirms that the spin-exchange is by far the strongest process in the system if $e$--$e$ losses are not present, as is the case in most recent proposals. The remaining inelastic dynamics is then limited to the \fs{$eg^+$} channel occurring in nuclear spin \fs{singlets}, which are suppressed in many-body experiments in our case of a large \fs{ferromagnetic} exchange interaction \fs{($V_{ex} > 0$)}.

With the large ratio of elastic to inelastic scattering length found in the inter-orbital interactions, a large part of \fs{phase diagrams of two-orbital physics analogous to that} discussed in \cite{Gorshkov2010a} appears accessible. Specifically, a \fs{ferromagnetic} coupling between the orbitals is realized, \fs{similar to condensed matter systems such as manganese oxide perovskites}.
The large absolute strength of the interaction term found in our work is desirable for the implementation of orbital magnetic phases, as tuning to lower values can be achieved by reducing the overlap between the $e$ and $g$ lattice potentials \cite{Daley2008-QuantComp}, which also further reduces inelastic collision rates accordingly.
The realization of an $^{173}$Yb lattice gas which exhibits two-orbital spin exchange, the elementary building block of orbital quantum magnetism, is therefore a fundamental step towards the realization of paradigmatic phases of matter and their SU($N$)-symmetric extensions.

\section{Methods}
\noindent
\textbf{Preparation of a two-component Fermi gas of $^{173}$Yb.} After Zeeman slowing on the \sz$\rightarrow^1$P$_1$ transition and cooling in a MOT on the \sz$\rightarrow^3$P$_1$ transition, approximately $10^7$ atoms of \textsuperscript{173}Yb ($I=5/2$) are loaded into a crossed optical dipole trap at 1064\,nm, where the desired spin mixture is prepared by a multiple-step optical pumping scheme. 
Evaporative cooling is carried out until the gas reaches Fermi degeneracy, with typically $N \simeq 1\times10^5$ atoms at $T/T_F = 0.25(5)$. The cloud is then loaded into a deep optical lattice operating at the state-independent (``magic'') wavelength of 759.354\,nm \cite{Barber2008-MagicLattice}.
Absorption imaging is performed using linearly polarized light, which minimizes the effect of the nuclear spin on the detection. Spin-resolved atom number detection is achieved by means of an optical Stern-Gerlach scheme, based on $\sigma^+$-polarized light, blue-detuned by $\simeq850$\,MHz from the MOT transition \cite{Taie2010,Stellmer2011} (see also Supplementary Information).

\vspace{2mm}
\noindent
\textbf{Interaction clock spectroscopy in a 3D lattice.} 
A two-component Fermi gas is loaded in 200\,ms into a 3D optical lattice with a typical mean depth of $\tilde{V}=43 E_r$. 
In order to probe the ultranarrow \sz$\rightarrow$\pz transition at 578\,nm, an IR diode laser at 1156\,nm is stabilized to a ULE cavity reference with a finesse $\mathcal{F} \simeq 10^5$. 
The stabilized laser light is then doubled in frequency by means of a SHG crystal placed in a bow-tie cavity. The clock excitation beam is overlapped with one of the lattice axes and is $\pi$-polarized along a uniform external B-field. Spectroscopy runs are performed to reference the absolute frequency in spin-polarized samples with low atom number $N \simeq 3 \times 10^4$ and a drift of the ULE cavity reference of $\simeq 2.5$\,Hz/min is compensated by applying a linearly varying frequency offset. 
Interaction spectroscopy runs are carried out by scanning over different state resonances in an alternating sequence, in order to prevent residual systematic drifts of the ULE cavity reference to affect the measured clock shifts. Each run is then separately referenced using the centre between the singly-occupied site resonances, which corresponds to the single-atom transition frequency in the absence of a magnetic field. A minimum absorption linewidth of 40\,Hz, limited by the laser frequency stability, was measured by spectroscopy on spin-polarized samples.

\vspace{2mm}
\noindent
\textbf{Dynamical equilibration of spin distributions in a 1D lattice.} Atoms are loaded into a $50\,E_r$ deep vertical optical lattice in 200\,ms and a  0.2\,ms clock transition $\pi$-pulse in a magnetic field of 20\,G is used to prepare
a balanced $\gu-\ed$ mixture. The spin dynamics is initiated by ramping the magnetic field to 1\,G in 200\,$\mu$s. After a variable holding time, the magnetic field is ramped back up and the spin state populations are independently detected after a $14\,$ms time-of-flight directly out of the lattice, by means of absorption imaging and optical Stern-Gerlach separation.

\vspace{2mm}
\noindent
\textbf{Determination of inelastic loss rates.}
Inelastic $e$--$e$ state losses are independently measured by applying a $\pi$-pulse to a gas loaded into the vertical optical lattice at zero magnetic field, exciting therefore an equal portion of both spin states.
After the residual ground state atoms are removed by a blast pulse resonant with the strong \sz$\rightarrow ^1$P$_1$ transition, the excited atomic population is monitored at varying hold time, by mapping it back to the ground state with a second $\pi$-pulse.
In order to address inelastic $g$--$e$ state losses, a mixture of $\gu$ and $\ed$ atoms is prepared by a $\pi$-pulse in a magnetic field of 20\,G. In separate measurement runs, the populations of the ground state and of the excited state are monitored at varying hold time.

\section{Additional information}
\noindent
Supplementary Information is available in the online version of the paper. Correspondence and requests for materials should be addressed to S.F.

\section{Acknowledgments}
\noindent
We gratefully acknowledge contributions by C. Schweizer, E. Davis and P. Ketterer during the construction of the experiment, and helpful discussions with A.M. Rey, M. Wall and A. Daley. This work was supported by the EU through the ERC Synergy Grant UQUAM and through the Marie Curie program (fellowship to P.C.D.G.).

\emph{Note added:} During the preparation of this manuscript, we became aware of related work on~$^{87}$Sr~\cite{Zhang2014-SUN-orbital-arxiv}.
\vspace{2\baselineskip}



\renewcommand{\thefigure}{S\arabic{figure}}
 \setcounter{figure}{0}
\renewcommand{\theequation}{S.\arabic{equation}}
 \setcounter{equation}{0}
 \renewcommand{\thesection}{S.\Roman{section}}
\setcounter{section}{0}
\renewcommand{\thetable}{S\arabic{table}}
 \setcounter{table}{0}

\newpage

\onecolumngrid
\begin{center}
\noindent\textbf{Supplementary Information:}
\bigskip

\noindent\textbf{\large{Observation of two-orbital spin-exchange interactions \\ with ultracold SU($N$)-symmetric fermions}}

\bigskip
F. Scazza$^{1,2}$, C. Hofrichter$^{1,2}$, M. H\"ofer$^{1,2}$, P. C. de Groot$^{1,2}$, I. Bloch$^{1,2}$ \& S. F\"olling$^{1,2} $
\vspace{0.1cm}

\small{$^1$ \emph{Fakult\"at f\"ur Physik, Ludwig-Maximilians-Universit\"at,\\ Schellingstrasse 4, 80799 M\"unchen, Germany}}

\small{$^2$ \emph{Max-Planck-Institut f\"ur Quantenoptik,\\ Hans-Kopfermann-Strasse 1, 85748 Garching, Germany}}
\end{center}
\bigskip

\section{Experimental sequence}

\begin{figure}[ht]
\begin{center}
\vspace{4mm}
\includegraphics[width=14cm]{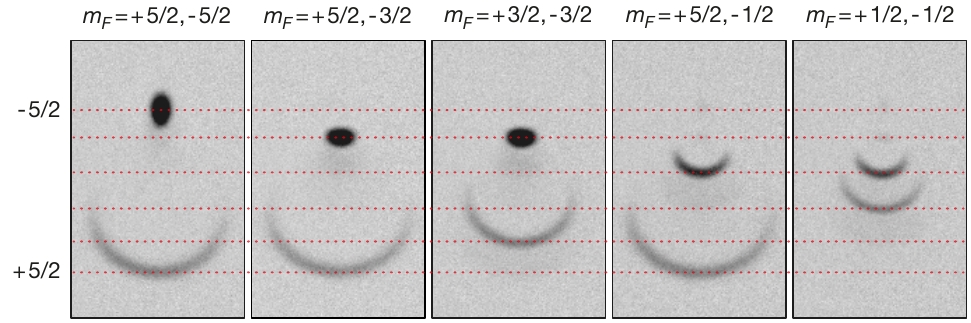}
\caption{\textbf{Time-of-flight absorption imaging of two-component Fermi gases with OSG.} Spin-resolved images of different spin mixtures which were used in the reported measurements. A 4\,ms OSG pulse is employed, with $30\,$mW light power and $80\,\mu$m waist at the atoms position.
}\label{fig:S1}
\end{center}
\end{figure}

Atoms of \textsuperscript{173}Yb ($I=5/2$) out of a thermal beam are first slowed on the strong \sz$\rightarrow^1$P$_1$ transition ($\Gamma/2 \pi = 29$\,MHz) at 399\,nm. The decelerated atoms are collected and cooled in a MOT, working on the narrow intercombination \sz$\rightarrow^3$P$_1$ transition ($\Gamma/2 \pi = 182$\,kHz) at 556\,nm. 
After a loading time of 8\,s, the power and the detuning from resonance of the MOT beams are decreased in 50\,ms, to let atoms cool down to a temperature $T\simeq$ 20\,$\mu$K. An optical dipole trap formed by two crossed beams at 1064\,nm is then superimposed to the MOT and approximately $10^7$ atoms are loaded into the trap in 200\,ms.
During the initial stage of the evaporation, the nuclear spin distribution is manipulated by means of optical pumping. A magnetic field of 20\,G is applied in order to split the different hyperfine transitions of the $^1$S$_0\rightarrow^3$P$_1$ $(F=5/2 \rightarrow F'=7/2)$ line. A $\sigma^+$-polarized and a $\sigma^-$-polarized light beam at 556\,nm are then used to independently address $m_F \rightarrow m_F\pm 1$ transitions and prepare the desired spin mixture by multiple light pulses. Clock transition spectroscopy runs are performed to reference the absolute frequency in completely spin-polarized samples with low atom number, in order to exclude any interaction-induced shifts. 
Evaporative cooling is carried out for 15\,s, typically reaching a temperature $T/T_F = 0.25(5)$ at $N \simeq 10^5$ atoms, and the cloud is then loaded into an optical lattice operating at the magic wavelength 759.354\,nm. 
Absorption imaging is performed using a linearly polarized light pulse resonant with the \sz$\rightarrow^1$P$_1$ transition.
In order to separately detect the population of the different nuclear spin states, an optical Stern-Gerlach (OSG) scheme is used: a $\sigma$-polarized 556\,nm light pulse with $\simeq 850\,$MHz blue detuning from the $^1$S$_0\rightarrow^3$P$_1$ ($F=5/2\rightarrow F'=7/2$) is shone on the atoms for $4\,$ms, with the maximum intensity gradient located at the atomic cloud centre. Time-of-flight images with the application of the optical Stern-Gerlach scheme are shown in Fig.~\ref{fig:S1} for various spin mixture preparations. 
The ground state SU($N$) symmetry is verified by monitoring the collisional stability of a two-component degenerate spin mixture in the dipole trap: no spin relaxation induced by spin-changing collisions is detected after up to 15\,s of holding time. 

\begin{figure}[!t]
\begin{center}
\includegraphics[width=10cm]{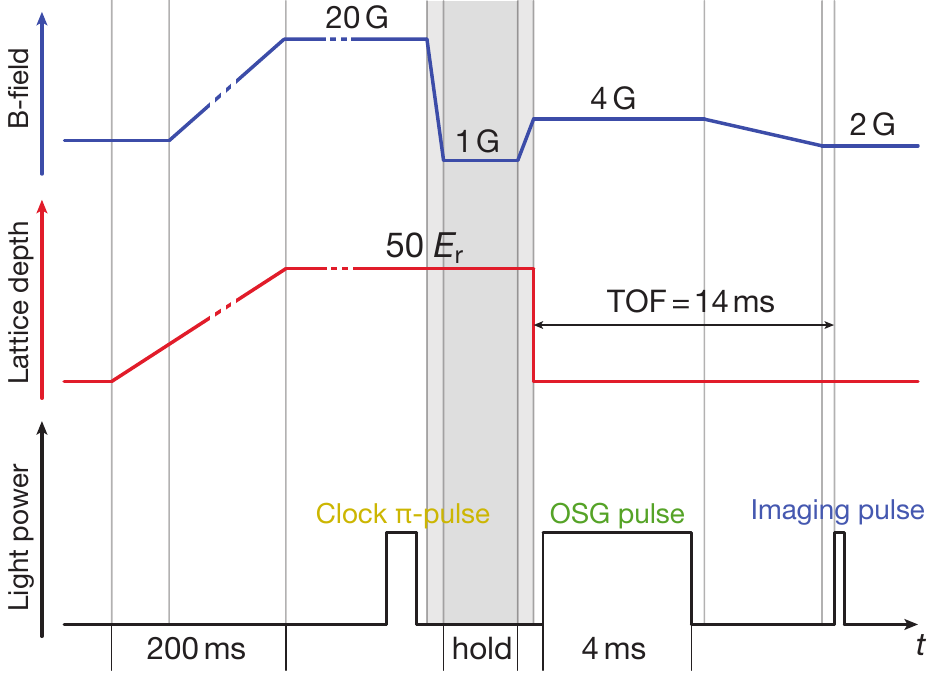}
\caption{\textbf{Experimental sequence for the observation of spin-exchange dynamics in a 1D optical lattice.}}
\label{fig:S2}
\end{center}
\end{figure}

The experimental sequence used for the observation of spin exchange in 2D ensembles is schematically illustrated in Fig.~\ref{fig:S2}. Atoms are loaded into the vertical optical lattice in $200\,$ms and a clock transition $\pi$-pulse is used to prepare a balanced $\gu$-$\ed$ mixture. The spin dynamics is initiated by ramping the magnetic field to 1\,G in $200\,\mu$s. After a variable holding time, the magnetic field is ramped back up and the relative occupation of the spin states is detected by time-of-flight OSG imaging directly out of the lattice. 

\section{Determination of exchange energy and SU($N$) symmetry}

As described in the main text, the transition to the lower eigenstate of Eq.~(2) exhibits a magnetic field dependence, which in a single-band model is given by Eq.~(3).
By fitting with Eq.~(3) we obtain $V$ and $\Vex$, and consequently $U_{eg}^+ - U_{gg}$ and $U_{eg}^- - U_{gg}$. A summary of results, corresponding only to the measurements of a $m_F=\pm 5/2$ mixture for three different lattice depths (shown in the inset of Fig.~3 in the main text), is given in Table~\ref{tab:S1}.

\begin{table}[htbp]
\begin{center}
		\begin{tabular}{| c | c | c | c |}
		\hline
		& & & \\
    \:$(V_x V_y V_z)^{1/3} \:[E_r]$ \:& \:\fs{$U_{eg}^- - U_{gg}$} [$h \cdot$kHz] \:&\: $\Vex$ [$h \cdot$kHz]     \:&\: \fs{$U_{eg}^+ - U_{gg}$} [$h \cdot$kHz] \: \\	
		& & & \\
		\hline
	  & & & \\
    $30$                        & $0.28 \pm 0.14$                       & \: \fs{$14.73 \pm 2.40$} \:                & $29.75 \pm 4.80$ \:\\ 
		& & & \\
		$33$                        & $0.33 \pm 0.05$                       & \: \fs{$16.16 \pm 0.83$} \:                & $32.65 \pm 1.65$ \:\\ 
		& & & \\
		$43$                        & $0.37 \pm 0.05$                       & \: \fs{$22.25 \pm 0.95$} \:                & $44.87 \pm 1.89$ \:\\
		& & & \\
    \hline
		\end{tabular}
\end{center}		
	\caption{Fit results for the $m_F=\pm 5/2$ mixture at different lattice depths.}
	\label{tab:S1}
\end{table}

\begin{figure}[hptb]
\begin{center}
\includegraphics[width=\textwidth]{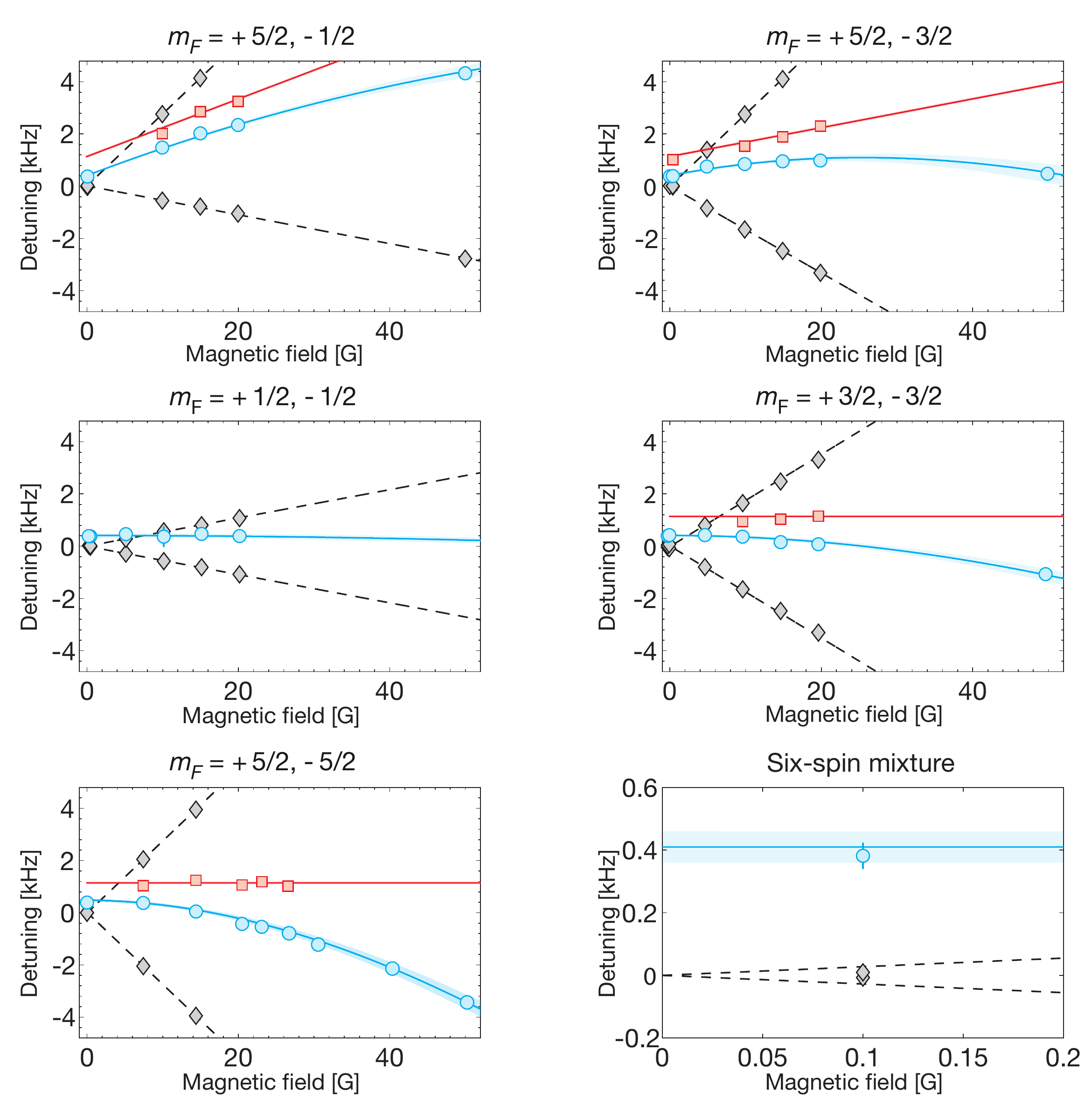}
\caption{\textbf{Resonance position shifts at varying magnetic field in different spin mixtures.} Grey diamonds mark singly-occupied site resonances, blue circles mark resonances to the lower eigenstate of Eq.~(2), and red squares mark $\ket{ee}$ state resonances. The lattice depth is $\tilde{V}=42.5 E_r$.
Solid blue lines are obtained through Eq.~(\ref{eq:eigenv-generalized}) with $V-U_{gg}$ and $\Vex$ obtained from fitting the $m_F=\pm 5/2$ mixture data. Solid red lines correspond to the value of $U_{ee}$ obtained through Eq.~(1) using $a_{ee}$ given in the main text. Shaded areas correspond to 95\% confidence intervals of the fits. Data point error bars represent the confidence intervals of the resonance position fits and are mostly hidden by data points.
} 
\label{fig:SUN_check2}
\end{center}
\end{figure}

\begin{figure}[htp] 
\begin{center}
\includegraphics[width=14cm]{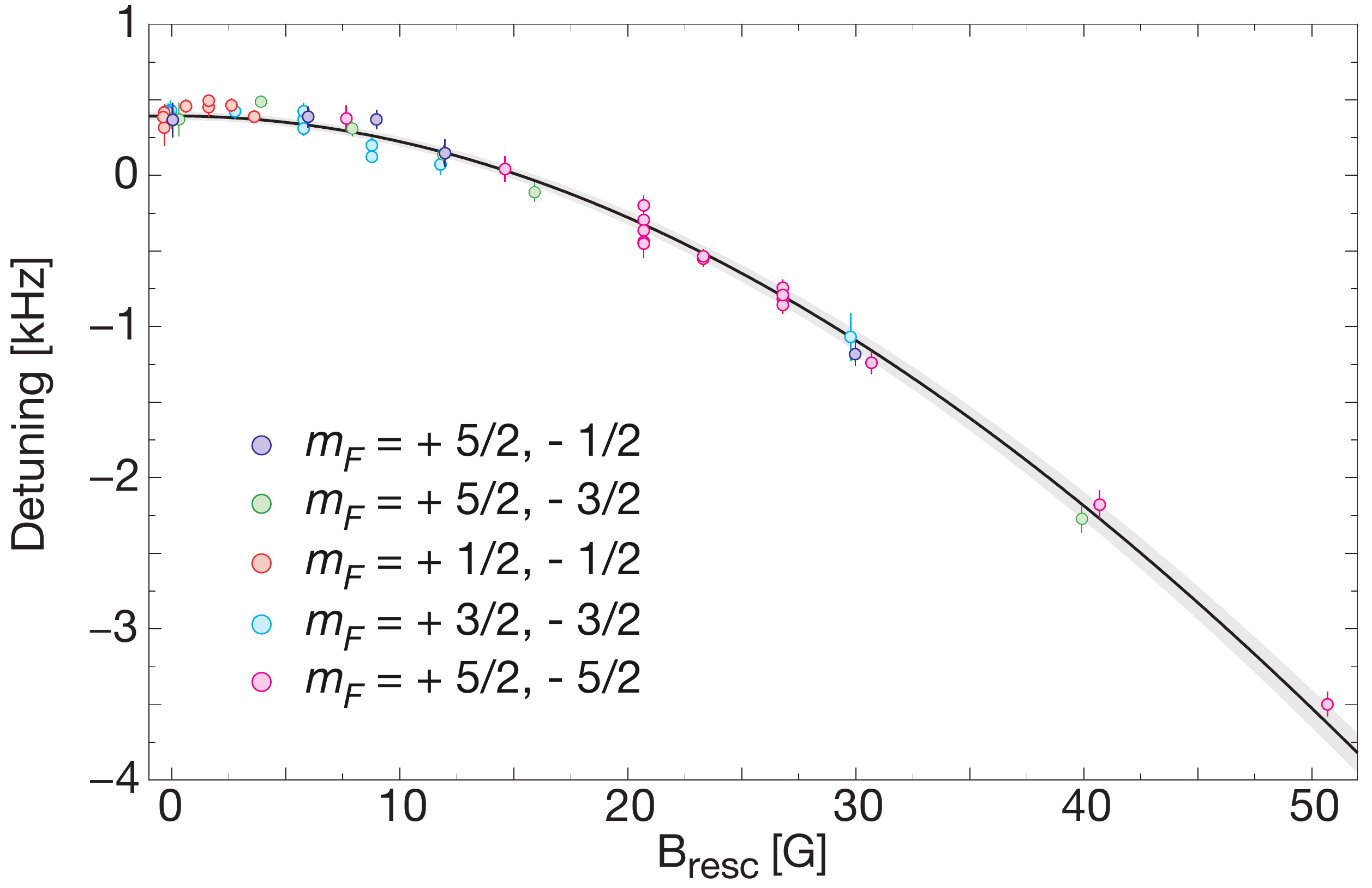}
\caption{\textbf{Shift of the transition to the lower eigenstate of Eq.~(2) in different spin mixtures.} The data shown here corresponds to lattice depths in a range between $\tilde{V}= 41 E_r$ and $\tilde{V}= 43 E_r$. The magnetic field axis has been rescaled to $B_{\text{resc}}=\frac{1}{5} (m_F - m'_F) B$ and the linear Zeeman shift of spin mixtures with $m'_F \neq -m_F$ has been subtracted.  
The black solid curve is obtained through a fit of all data points with Eq.~(\ref{eq:eigenv-generalized}), and the shaded area around the curve corresponds to the fit 95\% confidence interval. Data point error bars denote the confidence interval of the resonance position fits.}
\label{fig:SUN_check1}
\end{center}
\end{figure}

Data for the transitions to the lower eigenstate and to the \ket{ee} state in different spin mixtures are shown in Fig.~\ref{fig:SUN_check2}-\ref{fig:SUN_check1}.
The values of \fs{$a_{eg}^-$}, $a_{ee}$ and the respective uncertainties given in the main text are estimated through the data points shown here.
A very good agreement is found between parameter results associated to different spin mixtures, demonstrating the SU($N$)-symmetric nature of these interaction channels. 

A generalization of Eq.~(3) to arbitrary nuclear spin states $m_F$ and $m'_F$ is used in order to determine the resonance shifts in the curves of Fig.~\ref{fig:SUN_check2}-\ref{fig:SUN_check1}:
\begin{align}
h \cdot \Delta\nu(B) = \frac{1}{2}\, (m_F + m'_F)\,\delta g\, \mu_B B + \left(V - U_{gg}\right) - \sqrt{\Vex^2 + \left(\frac{1}{2} \left(m_F-m'_F\right) \delta g\, \mu_B B\right)^{\!\!2}}
\label{eq:eigenv-generalized}
\end{align}
For all blue curves in Fig.~\ref{fig:SUN_check2}, the $V-U_{gg}$ and $\Vex$ parameters obtained from the $m_F=\pm 5/2$ mixture data are being used. These are the ones with the smallest uncertainty amongst the fit results from all spin mixtures which are used in the experiment. All data sets are completely consistent with these fit results, demonstrating the SU($N$)-symmetric character of $U_{eg}^+$ as well, within our experimental uncertainty.
The solid red lines in Fig.~\ref{fig:SUN_check2} are determined using a single value of $U_{ee}$, obtained by inserting the estimated $a_{ee}$ given in the main text into Eq.~(1).   

\section{Multiband model for strong interactions}

In order to account for the large interaction strength \fs{$U_{eg}^+$} which we experimentally detect, it is necessary to take higher lattice bands into account to accurately estimate the interaction strength itself and the associated scattering length \fs{$a_{eg}^+$}. 

We performed a numerical diagonalization of the Hamiltonian including the lowest four energy bands of the lattice.
The total Hamiltonian is given by:
\begin{equation}
  \hat{H} = \hat{H}_{\mathrm{at},2} \otimes \mathbbm{1}
  + \mathbbm{1}\otimes \hat{H}_{\mathrm{lat},2}
  + \hat{U}\:,
  \label{eq:Htot}
\end{equation}
where the first Hilbert subspace refers to atom internal degrees of freedom (electronic and nuclear spin state) and the second Hilbert subspace refers to the motional degree of freedom (lattice vibrational states). 
The first term in the total Hamiltonian (\ref{eq:Htot}) accounts thus for the atom internal state, the second term for vibrational excitations and the third term for onsite contact interactions. The basis of completely symmetrized/anti-symmetrized two-particle states was used to numerically compute the Hamiltonian, where the interaction matrix elements were derived from computing the overlaps of onsite spatial wavefunctions of single-particle states. 
In order to remove bosonic states from the final solution an artificial large energy offset is added to all the symmetric states.
The scattering length \fs{$a_{eg}^-$} was fixed to the measured value, whereas the scattering length \fs{$a_{eg}^+$} was adjusted to make the lowest eigenenergy of (\ref{eq:Htot}) best reproduce the experimental data for the $m_F=\pm 5/2$ mixture. 
In our four-band numerical diagonalization, a non-regularized Dirac delta contact potential was used, and this is expected to significantly underestimate the eigenenergies with respect to their true value \cite{Mark2011}. 
In order to obtain accurate values of the scattering lengths from onsite interaction shifts a more precise treatment is required \cite{Mark2011, Mark2012, busch98a}, with the relevant models extended by the inclusion of the additional internal structure.

\bigskip
We also compared our data to the case of an infinite scattering length \fs{$a_{eg}^+$}, thereby assuming full ``fermionization'' of the onsite wavefunction in 1D. We use the band gap energy for \fs{$U_{eg}^+$}, which corresponds to the analytic result of the regularized treatment, and we use the fully fermionized onsite spatial wavefunctions instead of the non-interacting wavefunctions. The magnetic field dependence is then computed through the spatial overlaps between the fermionized eigenstates and the non-interacting eigenstates, which are respectively valid for the nuclear spin \fs{singlet} and \fs{triplet} subspaces. 
The calculated magnetic field dependence of the lower eigenenergy branch is still close to the measured data, therefore also implying a very large \fs{$a_{eg}^+ > a_{ho}$}.


\section{In-plane density model in a 1D optical lattice}

In order to estimate the density of the 2D trapped gas in each of the vertical optical lattice sites, we use a model which assumes conservation of total entropy and thermal equilibrium during the loading from the dipole trap into the lattice.
The profile of the Fermi gas in the dipole trap before the transfer into the lattice is modelled by a finite-temperature distribution obtained by local density approximation and valid in the limit of weak interaction.
The trap frequencies are determined experimentally by transferring momentum to the atom cloud and measuring the sloshing frequency in the dipole trap. For the trap configuration as it is at the transfer into the optical lattice we measure the following frequencies: $\omega_x=2\pi \cdot 8(1)\,$Hz, $\omega_y=2\pi \cdot 27(1)\,$Hz and $\omega_z=2\pi \times 212(5)\,$Hz.
After transfer into the lattice, the vertical degree of freedom is effectively frozen out by the lattice potential. The sum of entropies and atom numbers of 2D gases in the vertical lattice are then fitted to give the total entropy $S_0$ and atom number $N_0$ in the dipole trap, with the chemical potential $\mu_0$ and the temperature $T$ as free parameters:
\begin{equation}
	\begin{aligned}
		\begin{cases} S_0 = &\sum_{j} S_j^{2D}(\mu_j,T)\\
  	N_0 = &\sum_{j} N_j^{2D}(\mu_j,T)\:,\end{cases}
	\end{aligned}
\end{equation}
where $\mu_j= \mu_0 - \frac{1}{2} m \omega_z^2 z_j^2$ is the local chemical potential and $z_j$ is the $z$-coordinate of the center of the $j$-th vertical lattice site. An expression for the entropy of a Fermi gas as a function of $T$ and $\mu$ can be found for example in Ref. \cite{Kohl2006}.
The density profile in each lattice site is modelled then as a 2D finite-temperature Fermi-Dirac profile in the horizontal plane and as the Gaussian vibrational ground state in the vertical direction:
\begin{align}
n_j(x,y,z)=&\:s \frac{m k_B T}{2 \pi \hbar^2} \log\left[1 + \exp\!\left(\frac{\mu_j}{k_B T}\right) \exp\!\left(-\frac{m}{2 k_B T} \left(\omega_{x,\text{lat}}^2 x^2 + \omega_{y,\text{lat}}^2 y^2 \right)\right)\right]\notag\\
&\times \left(\frac{m \omega_{z,\text{lat}}}{\pi \hbar}\right)^{1/2} \exp\!\left(-\frac{m \omega_{z,\text{lat}}}{\hbar} \left(z - z_j \right)^2 \right)\:,
\end{align}
where $s$ is the number of spin components and $k_B$ is the Boltzmann constant.
Horizontal confinement frequencies in the lattice are measured to be $\omega_{x,\text{lat}}=\omega_{y,\text{lat}}=2\pi \cdot 37.5(10)\,$Hz and the vertical lattice band excitation frequency at a depth of $50\,E_r$ is $\omega_{z,\text{lat}}=2\pi \cdot 28.1(3)\,$kHz, leading to a typical mean density between $4 \cdot 10^{13}$ and $8 \cdot 10^{13}\,$ atoms/cm$^3$. These correspond to average in-plane Fermi energies between $E_F = h \cdot 2.1\,$kHz and $E_F = h \cdot 3.7\,$kHz.

\section{Spin-exchange dynamics and two-body losses}
\noindent
\subsection*{Spin-exchange dynamics in a 1D optical lattice}

\noindent
The difference in the symmetric and antisymmetric interaction strengths leads to an effective onsite magnetic interaction between two atoms in different electronic orbitals. For a more detailed discussion of the total Hamiltonian, see Ref. \cite{Gorshkov2010ab}. In a single-band description, the inter-orbital onsite interaction part of the Hubbard Hamiltonian can be written as 
\begin{align}
	H_{\text{int}} = V \sum\limits_i n_{ie} n_{ig} + \Vex \sum\limits_{imm'} c^\dagger_{igm} c^\dagger_{iem'} c_{igm'} c_{iem}\,,
\end{align}
where the operator $c^\dagger_{i\alpha m}$ creates an atom in the state $\ket{\alpha m}$ at site $i$ of the lattice, $\alpha=g,e$ and $m=\,\uparrow,\downarrow$\,, and $n_{i\alpha}=\sum_m c^\dagger_{i\alpha m} c_{i\alpha m}$.  
When the system is prepared in a superposition of the symmetric and antisymmetric states $\egp$ and $\egm$, the second term in this Hamiltonian is responsible for onsite spin-exchange oscillations.
For example, the superposition state
\[\ket{\psi (t=0)} =  \frac{1}{\sqrt{2}}\left(\egp + \egm\right) = \ket{eg\!\uparrow\downarrow}\]
evolves as 
\[\ket{\psi (t)} = \cos \left(\frac{\Vex}{\hbar} t\right) \ket{eg\!\uparrow\downarrow} + \sin \left(\frac{\Vex}{\hbar} t\right)\ket{eg\!\downarrow\uparrow}\:
\]
in the absence of magnetic fields. In the opposite regime of strong magnetic fields the exchange is energetically inhibited when the differential Zeeman shift between $\ket{e}$ and $\ket{g}$ states is larger than the exchange coupling strength.
 
A description based on binary interactions is still suitable when atoms are loaded into a 1D optical lattice. For atoms in different electronic orbitals, two scattering channels are available, characterized by different scattering lengths $a_{eg}^+$ and $a_{eg}^-$. The interaction couples therefore the $\ket{eg\!\downarrow\uparrow}$ and $\ket{eg\!\uparrow\downarrow}$ states. 
This is analogous to the description given in \cite{Stamper-Kurn2013} for single-orbital non-SU($N$)-symmetric spin-1 bosons. Recently, oscillating spin dynamics was observed also in a Fermi sea under specific initial conditions \cite{Krauser2014}, followed by a long-term spin relaxation dynamics \cite{Ebling2014}.

In our experiment, the gas is prepared in a $\gu$-$\ed$ balanced mixture with a $\pi$-pulse in a large field.
Strong $e$-$e$ pair lossy inelastic collisions are initially suppressed by Pauli blocking, as the atoms in the excited orbital initially are spin-polarized to $\ed$. As soon as a spin exchange occurs, an $\eu$ atom is present, which can collide with an $\ed$ atom without Pauli blocking. The population in the $\eu$ state is therefore strongly suppressed due to the $e-e$ loss process. As a consequence, this also strongly suppresses the build-up of coherence between the pair states involved in the spin-exchange process, which motivates the use of a simple classical rate equation based on spin-exchanging binary collisions, that we express in analogy to the description of inelastic binary collisions in terms of rate coefficients. The spin-exchange collisions themselves are slightly off-resonant, as there is a Zeeman shift difference of $\Delta_Z \approx +550$\,Hz between final and initial state a 1\,G field. This leads to a reduction of the rate, and can suppress the dynamics very strongly for higher fields. This also implies that the collisions are mode-changing with respect to the harmonic oscillator states in the transverse confinement of the trap. Finally, Pauli blocking effects of the processes due to occupied final states are not considered.

The spin-exchange process removes atoms from the initially prepared states $\gu$ and $\ed$ and transfers them to the initially unoccupied $\gd$ and $\eu$ states until equilibrium is reached. In order to include two-body loss processes, the rate equation for the density of atoms has also loss terms proportional to the product of the densities of the two components involved in the loss \cite{Stoof1988, Durr2009}.
Defining the relative state populations of the two electronic and spin states $P_{\alpha m}(t) = n_{\alpha m}(t)/n_0$, where $n_0$ is the density of the initial gas, the evolution can therefore be described using a simple system of coupled rate equations:

\begin{equation}
	\vspace{3mm}
		\begin{aligned}
		\dot{P}_{g\uparrow}(t)&=n_0 \gamma_{\text{ex}} \left(P_{e\uparrow}(t) P_{g\downarrow}(t) - P_{e\downarrow}(t) P_{g\uparrow}(t)\right) - n_0 \beta_{eg} P_{g\uparrow}(t) \left(P_{e\uparrow}(t) + P_{e\downarrow}(t)\right)\\
		\dot{P}_{g\downarrow}(t)&=n_0 \gamma_{\text{ex}} \left(P_{e\downarrow}(t) P_{g\uparrow}(t) - P_{e\uparrow}(t) P_{g\downarrow}(t)\right) - n_0 \beta_{eg} P_{g\downarrow}(t) \left(P_{e\uparrow}(t) + P_{e\downarrow}(t)\right)\\
		\dot{P}_{e\uparrow}(t)&=n_0 \gamma_{\text{ex}} \left(P_{e\downarrow}(t) P_{g\uparrow}(t) - P_{e\uparrow}(t) P_{g\downarrow}(t)\right) - n_0 \beta_{eg} P_{e\uparrow}(t) \left(P_{g\uparrow}(t) + P_{g\downarrow}(t)\right) - n_0 \beta_{ee} P_{e\uparrow}(t) P_{e\downarrow}(t)\\
		\dot{P}_{e\downarrow}(t)&=n_0\gamma_{\text{ex}} \left(P_{e\uparrow}(t) P_{g\downarrow}(t) - P_{e\downarrow}(t) P_{g\uparrow}(t)\right) - n_0 \beta_{eg} P_{e\downarrow}(t) \left(P_{g\uparrow}(t) + P_{g\downarrow}(t)\right) - n_0 \beta_{ee} P_{e\uparrow}(t) P_{e\downarrow}(t)	
		\end{aligned}
	\vspace{3mm}
	\end{equation}

\noindent
Three two-body rate coefficients are introduced: $\gamma_{\text{ex}}$ is the spin-exchange rate coefficient, $\beta_{ee}$ is the inelastic $e$--$e$ pair loss rate coefficient and $\beta_{eg}$ is the inelastic $e$--$g$ pair loss rate coefficient. All rate coefficients are independent upon the specific two-spin mixture, due to the collisional SU($N$) symmetry. The initial mean density of the gas $n_0$ is calculated as a weighted mean of single plane densities. 
We fit this model to the data with $\gamma_{\text{ex}}$, $\beta_{eg}$ and $P_{g\downarrow}(0)$ as free parameters. The sum $P_{e\downarrow}(0) + P_{g\downarrow}(0)$ is fixed to the fraction of $\ket{g\!\downarrow}$ atoms present without applying any excitation light, which is monitored to be constant during the experimental run. $P_{g\downarrow}(0)$ therefore accounts for imperfection of the clock $\pi$-pulse, which leaves typically 5-10\% of the \ket{\downarrow} atoms in the $\ket{g\!\downarrow}$ state. Similarly, $P_{g\uparrow}(0)$ is fixed to the fraction $\ket{g\!\uparrow}$ atoms in the gas in the absence of excitation light.
A 2-dimensional fitting procedure is used: the measured relative spin state population is fitted with $P_{g\uparrow}(t)/(P_{g\uparrow}(t) + P_{g\downarrow}(t))$ and the total detected ground state atom number is fitted with $N_{0g}\left(P_{g\uparrow}(t) + P_{g\downarrow}(t)\right)$, where $N_{0g}$ is the number of ground state atoms for zero hold time.

The absolute values of the rate coefficients in the $e$--$g$ mixture are thus based on a simple model for the effective densities, neglecting much of the complex many-body evolution in the presence of spin exchange, strong interactions and strong $e$--$e$ loss. Also the fact that the spin exchange is additionally affected by being off-resonant and therefore mode-changing in the 1\,G offset field is not taken into account.
\bigskip

\subsection*{Characterization of inelastic losses}

\noindent
As described in the Methods section of the main text, inelastic $e$--$e$ losses are measured separately, in order to fix the value of $\beta_{ee}$. The loss rate coefficient is obtained by fitting $P_{e\uparrow}(t) + P_{e\downarrow}(t)$ to the excited population data, using our rate equation model.\\  
In order to increase sensitivity to $e$--$g$ collisional loss, the spin-exchange was strongly reduced by holding the atoms at high magnetic field.
A mixture of $\gu$ and $\ed$ atoms is prepared by a $\pi$-pulse in a magnetic field of 20\,G and the total population of the ground state is measured after a varying hold time. In a separate data set the population of the excited state is also measured by mapping it back to the ground state using a second $\pi$-pulse. The loss rate coefficient is obtained by a combined fit of $P_{g\uparrow}(t) + P_{g\downarrow}(t)$ to the ground population data and $P_{e\uparrow}(t) + P_{e\downarrow}(t)$ to the excited population data, with the same rate equation model. This fit of $\beta_{eg}$ yields a value in good agreement with the ones obtained by fitting this parameter together with $\gamma_{\text{ex}}$ to the measurements in a 1\,G field, as explained above. These $e$--$g$ pair losses involve both symmetric and antisymmetric inelastic scattering channels, including scattering of particles in the triplet states with same nuclear spin orientation, and the reported $\beta_{eg}$ rate is therefore a combined effective loss rate. 

\begin{figure}[bth] 
\begin{center}
\includegraphics[width=11cm]{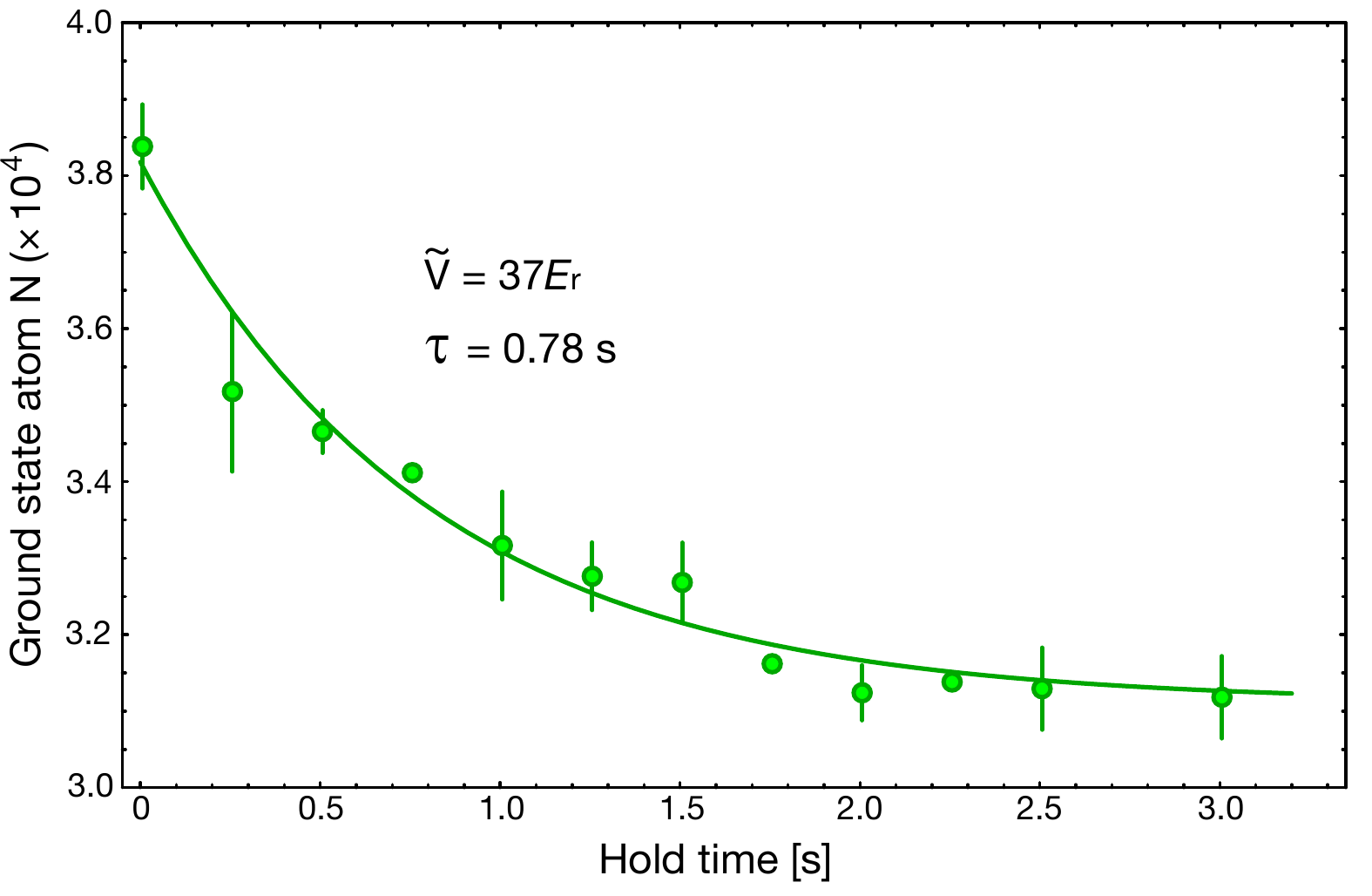}
\caption{\textbf{On-site atom losses after excitation to the \fs{$\egm$\xspace} state in a deep 3D lattice.} The ground state atom population is monitored after excitation with a $\pi$-pulse, resonant with the transition to the lower-lying two-particle eigenstate. The magnetic field is set to 7\,G and the mean lattice depth is $\tilde{V}=37 E_r$. 
The ground state atom number prior to the excitation pulse is $N = 4.7 \times 10^4$, i.e. $\simeq 0.9 \times 10^4$ atoms are excited by the $\pi$-pulse. The decay stops when nearly all atoms on doubly occupied sites have been lost. The solid line is an exponential fit, which yields a lifetime of $\tau=0.78$\,s, and the error bars denote standard deviation of the mean obtained by averaging typically 3-4 measurements.}
\label{fig:egPlusDecay}
\end{center}
\end{figure}

In order to independently determine the two-body loss rate of the \fs{$\egm$} pairs, thereby separate the symmetric and antisymmetric loss channels, the decay of doubly occupied sites in a 3D lattice is also characterized. For this, a $m_F=\pm 5/2$ spin mixture is loaded into a deep 3D lattice and excited to the \fs{$\egm$} state in a 7\,G magnetic field with a $\pi$-pulse.

The decay is characterized by monitoring the ground state atom number after a hold time in the lattice. The measured lifetime $\tau=0.78$\,s (see Fig.~\ref{fig:egPlusDecay}) is equivalent to a two-body loss rate \fs{$\beta_{eg}^-= 3 \times 10^{-15}$\,cm$^3$/s}, corresponding in turn to an imaginary part of the scattering length \fs{$\eta_{eg}^- = -0.006\,a_0$}. 
The relation between the two-body loss rate \fs{$\beta_{eg}^-$} and the onsite atom lifetime $\tau$ is, in a single-band model \cite{Garcia-Ripoll2009}:
\begin{equation}
  \frac{1}{\tau}= \fs{\beta_{eg}^-} \: \int \! d^3r\, w_0^4(\textbf{r})
\end{equation}
The loss rate can then be related to the imaginary part \fs{$\text{Im}(a_{eg}^-)=\eta_{eg}^-$} of the scattering length \cite{Mott1965, Bohn1997}:
\begin{equation}
	\fs{\beta_{eg}^- = - \frac{8 \pi \hbar}{m} \eta_{eg}^-} 
\end{equation}
\noindent
The total atom loss after a long hold time matches the amount of atoms which had been initially excited by the $\pi$-pulse, indicating a loss of atoms on the doubly occupied sites. 
The rate \fs{$\beta_{eg}^-$} is given here as an upper bound, as the rate of tunnelling events for atoms in a cubic lattice at the adopted depth becomes non-negligible at these time scales, enabling losses from the other collision channels.

\end{document}